\documentstyle[aps,prl,multicol,epsf]{revtex}
\def\eps{\epsilon}

\def\ket#1{| #1\rangle}

\def\lemma#1{\noindent{\bf Lemma$~#1$} \qquad}
\def\theorem#1{\noindent{\bf Theorem$~#1$} \qquad}

\def\proof{\noindent{\bf Proof} \qquad}

\def\R{\hbox{\rm I \kern-5pt R}}

\def\Tr{{\rm{Tr}}}

\title{Optimal Entanglement Enhancement for Mixed States}
\author{ Adrian Kent${}^1$, Noah Linden${}^2$ and Serge Massar${}^3$}
\address{${}^1$ Department of Applied Mathematics and Theoretical Physics,
University of Cambridge,\\ 
Silver Street, Cambridge CB3 9EW, UK\\
$^2$Isaac Newton Institute for Mathematical
Sciences, Cambridge, CB3 0EH, UK\\%
${}^3$ Service de Physique Th\'eorique,  
 Universit\'e Libre de Bruxelles, 
 CP 225, Bvd du Triomphe,       
 B1050 Bruxelles, Belgium. 
}
\date{4 February 1999}

\begin{document}
\maketitle
\begin{abstract}
We consider the actions of protocols involving local quantum 
operations and classical communication (LQCC) on a single system consisting  
of two
separated qubits.  We give a complete description of the orbits
of the space of states under LQCC and characterise the representatives
with maximal entanglement of formation.  We thus obtain
a LQCC entanglement concentration protocol for a single 
given state (pure or mixed) of two qubits which is optimal in the
sense that the protocol produces, with non-zero probability, a
state of maximal possible entanglement of formation.  
This defines a new entanglement measure, 
the {\it maximum extractable entanglement}.

\end{abstract}


\begin{multicols}{2}

\section{introduction}\label{introduction}

Entanglement is a basic
quantum communication resource which
can usefully be manipulated 
to suit particular tasks \cite{BBPS,BBPSSW}. 
In this paper we investigate the manipulation of a single
entangled mixed states comprising two separated 
single qubit subsystems. 
We consider two parties, Alice and Bob, who each control one 
subsystem, and who are restricted to carrying out  
local quantum operations and classical communication (LQCC). 
Specifically the quantum operations Alice and Bob are allowed to
perform are local unitary transformations and local filtrations.
The restriction to local quantum operations ensures that
entanglement is indeed treated as a resource: if non-local quantum 
operations were allowed, Alice and Bob could 
create entanglement between them from initially non-entangled states. 

The interest of this problem is that
since any real-world quantum communication channel will be imperfect,
even if Alice could create perfect maximally entangled 
states, she would never be able to share such states with
Bob simply by sending one subsystem through the channel.   
So it is natural to ask whether Alice and Bob can use LQCC to  
obtain states with better entanglement from imperfectly
entangled states.  

Various entanglement
purification protocols have been suggested. 
If Alice and Bob share a number of copies of an
imperfectly entangled known pure state, they can
obtain a number of maximally entangled states by carrying out 
operations on each state individually or by collective 
operations on a number of shared states \cite{BBPS}.
The collective algorithm has a higher asymptotic 
yield of maximally entangled states in the limit in which
the number of shared states tends to infinity.  
Efficient collective algorithms which give a non-zero asymptotic
yield of maximally entangled states from entangled mixed states
have also been described \cite{BBPSSW}. 

In practice, though, the number of states will always be finite, 
and Alice and Bob will effectively share a single entangled state 
of two subsystems whose state spaces are finite dimensional.  
For this and other reasons --- for example, Alice and Bob might
actually have only one copy of an entangled state of some 
simple system, or it may be technologically difficult to 
implement collective operations --- 
it is interesting to see what Alice and Bob can
achieve by gambling with the entanglement of a single
state.  That is, we would like to know how far the entanglement
of single states could be increased by LQCC if
the outcomes of Alice and Bob's local measurements were 
favourable.  The Procrustean
algorithm of \cite{BBPS} 
provides an answer to this question in the case of pure states.
Here we answer the 
question for two qubit mixed states, and in the process
illustrate a general approach to the problem based on
identifying quantities invariant under LQCC.

Though most mixed state entanglement distillation protocols
discussed so far involve collective operations on many 
states, it has been established that there exist
entangled mixed states for which single-state LQCC protocols can 
increase entanglement \cite{G2,H3}.
Conversely, it is known that there exist entangled 
mixed states, including the important case of the Werner
states, for which no single-state LQCC protocol can increase 
entanglement \cite{LMP,K,H4}. 

We give here a complete description of the effect of LQCC on 
entanglement of a single copy of an arbitrary mixed state
$\rho$ of two qubits. 
It has been shown previously that if Alice and Bob's local density
matrices are completely random, they cannot increase the Entanglement
of Formation (EOF) by LQCC\cite{LMP,K,H4}. 
Here we show that if the local density
matrices are not random and if the EOF is non-vanishing, Alice and Bob
can always increase the EOF. Moreover we construct a procedure that
maximises the EOF of the final state. This procedure, which is unique
up to local unitary transformations, leaves Alice and Bob with
completely random local density matrices.

\section{Main results}

Throughout, the states considered are those of a single system
comprising two separate single qubit subsystems.  
We use the following facts.\cite{LMP}
\begin{enumerate} 
\item 
The LQCC protocols we consider map the state $\rho$ to states of the form
\begin{equation}\label{image}
\rho' = \frac{ A \otimes B \rho A^{\dagger} \otimes B^{\dagger} }{
\Tr ( A \otimes B \rho A^{\dagger} \otimes B^{\dagger} ) } \ , 
\label{finalform}
\end{equation}
where $A$ and $B$ are arbitrary operators that act on Alice and Bob's
Hilbert space respectively.
The only condition they must obey is $A^\dagger A \leq I_2$, $B^\dagger
B \leq I_2$. 
The protocol succeeds with 
probability $\Tr ( A \otimes B \rho A^{\dagger} \otimes B^{\dagger} )
$.
We need not consider the most general local protocols in which the
final state consists of mixtures of states of the form
eq. (\ref{finalform}) since mixing decreases the EOF.

The operators $A$ and $B$ can be written as
\begin{equation}\label{one} 
A \otimes B = U_A f^{\mu,a,{\bf m}} \otimes 
U_B f^{\nu, b, {\bf n}} \, , 
\end{equation}
where $U_A , U_B$ are unitary and the filtrations $f$ are defined by 
\begin{equation}  
f^{\mu,a,{\bf m}} = 
\mu (I_2 + a {\bf m}. {\bf \sigma} )\ \hbox{\rm and}\ f^{\nu,b,{\bf n}} = 
\nu (I_2 + b {\bf n}. {\bf \sigma} )\, .
\label{ff}
\end{equation}
Here $\mu,\nu , a ,b $ are real numbers, $I_n$ denotes the identity
operator in $n$ dimensions, and the vector ${\bf \sigma} = \{ \sigma_1 , 
\sigma_2 , \sigma_3 \}$ has the Pauli matrices as components. 
We can also write these operators as $A = U_A F_A U'_A $,
where $F_A$ takes the form $ \left(\matrix{ \alpha_1  &  0   \cr 0 &
\alpha_2 } \right)$ 
with the $\alpha_i$ real, $0\leq \alpha_i \leq 1$ and $U_A , U'_A$
unitary; similarly $B = U_B F_B U'_B$.  
We can thus write any non-trivial LQCC (i.e. any LQCC which is not
the zero map) in the form
\begin{equation}\label{three}
\gamma U_A \left(\matrix{ 1  &  0   \cr 0 &
\alpha_A } \right) U'_A \otimes
U_B \left(\matrix{ 1  &  0   \cr 0 &
\alpha_B } \right) U'_B \, ,
\end{equation}
where $\gamma$ is a scale factor in the range $0 < \gamma \leq 1$
and $0 \leq \alpha_A , \alpha_B \leq 1$. 

\item  
The entanglement of formation (or EOF) of a pure state
$|\psi\rangle$ is defined as $E(\psi) = - Tr \rho_A \ln \rho_A = 
- Tr \rho_B \ln \rho_B$ where $\rho_A= Tr_B |\psi\rangle\langle \psi
|$, $\rho_B= Tr_A |\psi\rangle\langle \psi
|$ are the local density matrices seen by Alice and Bob. For a mixed
state the EOF is defined as\cite{BDSW}:
$E(\rho) = \min \sum_i p_i E(\psi_i)$ where the minimum is taken over
all decompositions of $\rho$ into pure states $\rho = \sum_i p_i
|\psi_i\rangle \langle \psi_i |$.

In the case of a mixed state comprised of two single qubit subsystems,
Wootters\cite{W1} has given an explicit formula for $E(\rho)$,
verifying an earlier conjecture of Hill and Wootters\cite{HW1}.
Let $\tilde{\rho} = \sigma_2 \otimes \sigma_2 
\rho^* \sigma_2 \otimes \sigma_2 $.
Call $\lambda_i$ the positive square roots of the eigenvalues
of the matrix $\rho \tilde{\rho }$ written in decreasing 
order. Define the concurrence by 
\begin{equation}
C ( \rho ) = \max ( 0, \lambda_1 - \lambda_2 - \lambda_3 - \lambda_4 )
\ .
\end{equation}
Then the EOF of $\rho$ is 
\begin{equation} 
E ( \rho ) = H ( \frac { 1 + \sqrt{ 1 - C^2 ( \rho ) }}{2} ) \, , 
\end{equation}
where 
$H ( p ) = - p \log_2 p - (1-p) \log_2 (1-p)$.

\item
Consider a general density matrix $\rho$ of two qubits. It can be
written as
\begin{equation}
\rho = \frac{1}{4}  ( I_4 + \alpha . \sigma \otimes I_2 
+ I_2 \otimes \beta . \sigma + R_{ij} \sigma_i \otimes \sigma_j ) \ .
\label{abc}
\end{equation}
In \cite{LMP} it was shown that under LQCC of the form eq. (\ref{one})
the positive square roots of the eigenvalues
of the matrix $\rho \tilde{\rho }$ transform as
\begin{equation}
\lambda_i \to \lambda'_i = \frac { \mu^2 \nu^2 ( 1 - a^2 ) (1 - b^2 ) }{ 
t(\rho; \mu, a , {\bf m}; \nu, b , {\bf n} ) } \lambda_i 
\label{lambda}
\end{equation}
where $t$ is the probability that the LQCC 
succeeded
\begin{eqnarray} 
&t&(\rho; \mu, a , {\bf m}; \nu, b , {\bf n}) =\nonumber\\
& &\quad \mu^2\nu^2\big[ (1 + a^2)(1+b^2) +
2 a(1+b^2) {\bf n} \cdot {\bf \alpha} + \nonumber\\ 
& &\qquad 2b(1+a^2) {\bf m}  \cdot  {\bf \beta} 
+ 4 ab R_{ij} 
n_i m_j\big].
\label{t}
\end{eqnarray}
Thus the concurrence also transforms as
\begin{equation}
C(\rho') = \frac { \mu^2 \nu^2 ( 1 - a^2 ) (1 - b^2 ) }{ 
t(\rho; \mu, a , {\bf m}; \nu, b , {\bf n} ) } C(\rho) \, .
\label{CCC}
\end{equation}
It follows from eq. (\ref{lambda}) that the ratios $\lambda_i /
\lambda_j$ are invariant under LQCC. We add here the necessary
qualification that the LQCC must be invertible.
\end{enumerate}

Now our argument runs as follows.  
We consider states $\rho$ which have non-zero EOF and which are not
Bell diagonal (recall that a state is Bell diagonal if all its
eigenvectors are maximally entangled; equivalently it satisfies
$tr_A(\rho)= tr_B(\rho)= \frac{1}{2} 
I_2$, ie. $\alpha=\beta=0$ in the expression
for $\rho$ given in eq. (\ref{abc})). We show in Theorem 1 that there
is 
an LQCC protocol which increases
the EOF of $\rho$ with non-zero probability.
We show further in Theorem 3 that this process can be iterated to obtain
 an 
LQCC protocol which,  with non-zero probability, maps $\rho$ to a 
Bell diagonal state with maximal EOF. 
In Theorem 4 we show that this is the unique optimal protocol  
up to local unitary rotations. 

\theorem{1}  Let $\rho$ be a density matrix
of a state with non-zero EOF  written as in eq. (\ref{abc}).
If $\alpha$ or $\beta$ are non-zero, then
there is an invertible LQCC $A \otimes B$ 
mapping $\rho$ with non-zero probability to a density matrix 
$\rho'$ with higher EOF than $\rho$.

\proof  
For small $a$ and $b$, eq. (\ref{CCC}) takes the form
\begin{equation}
C(\rho') \simeq \frac { 1 }{ 1 +  2a {\bf 
m}. \alpha + 2b 
 {\bf n} \beta} C(\rho) \, .
\label{CCC2}
\end{equation}
Hence if $\alpha$ or $\beta$ are non-zero and if $C( \rho )$ is 
non-zero we can always find an 
LQCC which,  with non-zero probability, increases
the EOF, by choosing appropriately
small $a$ and $b$ and suitable ${\bf m}$ and ${\bf n}$.  

\vskip 10pt

We now need a technical lemma about the topology of the space $R$ of
LQCC operations which do not decrease the EOF of a given $\rho$. The
result, namely that $R$ is compact, is needed in Theorem 3.

\lemma{2}  Let $\rho$ have non-zero EOF.  
There exists a positive bound $\delta (\rho )$ such that
if the state $\rho'$ has greater EOF than $\rho$ and can be obtained from
$\rho$ with non-zero probability by LQCC, then  there exists some LQCC
from which $\rho'$ can be obtained 
from $\rho$ with probability greater than $\delta ( \rho )$.  
Furthermore let $R$ be the space of LQCC which succeed with non-zero
probability in producing a density matrix with EOF greater than or
equal to that of $\rho$. Then $R$ is compact.

\proof  
Fix $\rho$. 
If we write $A \otimes B$ in the form (\ref{three}), $\rho'$ is 
independent of the scale factor $\gamma$, 
so that any $\rho'$ obtainable from $\rho$
can be obtained by a {\it normalised} LQCC, taking the form
(\ref{three}) with $\gamma =1$.  
For $\eps > 0$, define $S_{\epsilon}$ to 
be the set of normalised LQCC of the form (\ref{three}) 
with $\min\{ \alpha_A , \alpha_B \} = \eps$.
Let $
E_{\eps} $ be the maximum EOF of any density matrix 
$\rho'$ obtained from $\rho$ by the action (\ref{image}) for 
some $A \otimes B$ in $S_{\eps}$.  
Since $ 
E_{\eps}$ is continuous in $\eps$ and tends to zero
as $\eps$ tends to zero, there is some positive $\eps_0 $ 
such that $ 
E_{\eps} $ is less than or equal to the EOF of $\rho$ for
$\eps \leq \eps_0 $ and such that $\eps_0 (\rho )$ is 
maximal with this property. 
Let $T_{\eps_0}$ be the union for ${1\geq \eps\geq \eps_0}$ of
$S_{\eps}$.  
Now if a non-trivial LQCC $A \otimes B$ annihilates $\rho$, i.e. 
$ A \otimes B \rho A^{\dagger} \otimes B^{\dagger} = 0 $, 
then $A \otimes B \ket { \psi_i } = 0$ for all $i$ (where 
$\ket { \psi_i }$ are the eigenvectors of $\rho$ with non-zero
eigenvalue). Thus either
$A$ or $B$ must be a rank one projector up to a
scale factor.   Hence no LQCC in $T_{\eps_0}$ can annihilate
$\rho$.  Also $ 
T_{\eps_0}$ is compact.  So  the probability 
$ \Tr ( A \otimes B \rho A^{\dagger} \otimes B^{\dagger} )  $ of 
obtaining $\rho'$ from $\rho$ via the LQCC is non-zero 
everywhere in $T_{\eps_0}$ and attains a non-zero lower bound
$\delta(\rho)$
on the set.  This is a lower bound for all LQCC increasing 
the EOF of $\rho$, since no LQCC outside $
T_{\eps_0}$ does. The compactness of $R$ follows since it is a closed
subset of $T_{\eps_0}$.

\vskip 10pt

\theorem{3} Let $\rho$ written as in eq. (\ref{abc}) be a density
matrix with non-zero EOF. 
If $\alpha$ or $\beta$ are non-zero, then
there exists  an invertible LQCC 
which, with non-zero probability, maps  $\rho$ to a Bell diagonal density 
matrix
$\rho'$ which has the maximum EOF of any density matrix obtainable
from $\rho$ by LQCC.

\proof
Since by Lemma 2 the space of normalised LQCC which do not decrease 
the EOF of $\rho$
is compact, and the EOF is a 
continuous function, the lowest upper bound on the attainable
EOF is attained by some LQCC. 
The corresponding density matrix $\rho'$ must have  
$\alpha'=\beta'=0$, otherwise, by Theorem 1, its EOF could be increased.

\vskip 10pt

\theorem{4}  Let $\rho$ be the density matrix of a state with non-zero
EOF.  Then the Bell diagonal state $\rho'$
 which can be obtained from $\rho$ by LQCC is unique up to
local unitary transformations. This $\rho'$ has maximal possible EOF.

\proof   
We start by calculating the positive square roots $\lambda_i$ of the
eigenvalues of the matrix $\rho\tilde\rho$. We order them as
$\lambda_1 \geq \lambda_2 \geq\lambda_3 \geq\lambda_4$. 
The ratios $\frac{ \lambda_i }{ \lambda_j }$ are invariant
under the actions of invertible LQCC, see eq. (\ref{lambda}). 
We characterise these ratios by the three numbers $c_i =  
\lambda_i /  \lambda_1$, $i=2,3,4$. 

{}From Theorem 3 we know that $\rho$ can be brought to Bell diagonal
form by LQCC. We shall now show that the Bell diagonal form is
uniquely specified, up to local unitary transformations, by the ratios
$c_i$.
To this end consider a Bell diagonal state
$ \rho_{R} = \frac{1}{4} ( I_4 + R_{ij} \sigma_i \otimes \sigma_j ) $
with positive EOF. 
Local unitary operations $U_A \otimes U_B$ transform $ \rho_{R}$ 
to $ \rho_{R'} = \frac{1}{4} ( I_4 + R'_{ij} \sigma_i \otimes 
\sigma_j ) $, where $R' = ( O_1 )^T R (O_2 )$ for some
elements $O_1$ and $O_2$ of ${\rm SO}(3)$: any pair of $O_i$
can be produced by suitable choices of $U_A , U_B$. 
By using a singular value decomposition\cite{eglewis} of $R$, 
we can find orthogonal $O_i$ such that $R'$ is diagonal, so 
we can find
local unitary operations mapping $\rho_R$  to the form 
\begin{equation}
\rho_{r_1 , r_2 , r_3} =
\frac{1}{4} ( I_4 + \sum_{i=1}^3 r_i \sigma_i \otimes \sigma_i ) \, , 
\label{BD}
\end{equation} 
with all $r_i$ having the same sign and with $r_1 \leq r_2 \leq r_3$.

Now $\rho_{r_1 , r_2 , r_3} = \tilde{\rho}_{r_1 , r_2 , r_3}$,
hence the eigenvalues of $\rho_{r_1 , r_2 , r_3}$ are equal to the
$\lambda_i$. These eigenvalues are
$ \frac{1}{4} (1 - r_1 - r_2 - r_3 ), \frac{1}{4} (1 + r_1 + r_2 - r_3
) , \frac{1}{4} (1 + r_2 + r_3 - r_1), \frac{1}{4} (1 + r_3 + r_1 -
r_2 )$.  
Since $\rho_{r_1 , r_2 , r_3}$ is assumed to be entangled, the $r_i$
are all less then or equal to zero. (This may be verified by
checking that when the $r_i$ are all positive the concurrence
vanishes). 
We can
now express the ratio's $c_i$ in terms of the $r_i$. For instance
$c_2 = ( 1 + r_2 + r_3 - r_1) / (1 - r_1 - r_2 - r_3 )$. It is
straightforward to verify that the $r_i$ can be uniquely expressed in
terms of the $c_i$ by inverting these equations. Therefore the Bell
diagonal state of the form eq.(\ref{BD}) to which $\rho$ can be
brought is unique.

\vskip10pt
\section{Conclusions}

We have shown that any entangled state $\rho$ of two qubits
whose local density
matrices are not completely random can be brought by LQCC to a unique
(up to local unitary transformations) Bell diagonal state. No other
LQCC can bring $\rho$ to a state with more entanglement. 
To obtain an explicit expression for this optimal protocol, one should 
write explicitly the conditions that the density matrix $\rho'$
obtained from has completely random local density matrices $Tr_A \rho' 
= Tr_B \rho' = I_2$. We have shown that these equations have a unique
solution for the coefficients $a,{\bf m},b ,{\bf n}$ of the
filtrations
$f^{\mu,a,{\bf m}}, f^{\nu, b, {\bf n}}$ in eqs. (\ref{one},\ref{ff}).

Our optimal protocol 
should be compared to the Procrustean algorithm for 
concentrating pure state entanglement of \cite{BBPS} which
brings a non maximally entangled pure state to a maximally entangled
pure state by LQCC. 
The main difference between the two methods is that 
the optimal mixed state
protocol generally requires Alice and Bob to carry out
different filtrations and then tell each other whether the filtrations have
succeeded.  Only if both succeed do they obtain (and know that they
have) the state with maximum extractable entanglement. 
The Procrustean algorithm  on the other hand can be realised without
classical communication between Alice and Bob, or with only Alice
carrying out the filtration and communicating the result to Bob.

In \cite{LMP} it was noted that the ratios $c_i
= \lambda_i/\lambda_1$ are invariant under invertible LQCC. 
The argument used in proving Theorem 4 also shows that for entangled
states they consitute an exhaustive set. Indeed we can bring any
entangled $\rho$ to the form eq. (\ref{BD}) which is characterised by
three parameters $r_i$ and they are in one to one correspondence with
the $c_i$. This gives a characterisation of locally equivalent
entangled density matrices.

Our method also introduces an interesting combination of these
invariants: the maximal extractable entanglement of a density
matrix. This quantity provides a new characterisation of the
entanglement of a state. It has the important
property that it decreases under
mixing (this follows from the convexity of the EOF\cite{BDSW}).

\noindent{\bf Acknowledgments}
We are very grateful to Sandu Popescu for several helpful 
discussions.  Part of this work was caried out at the 1998 
Elsag-Bailey - I.S.I. Foundation research meeting on quantum
computation and at the 1998 workshop of the Benasque Center for Physics.
AK thanks the Royal Society for financial support. SM is a 
chercheur qualifi\'e  du FNRS.

\end{multicols}

\end{document}